\documentclass{jsp-book9x6}
\usepackage{jsp-book-har} 
\usepackage{subfigure,bm}
\usepackage[T1]{fontenc}

\begin{document}

\chapter[Gyroid ferromagnetic nanostructures in 3D magnonics] {Gyroid ferromagnetic nanostructures in 3D magnonics\label{3d_magn_crys}}

\chapauth{Mateusz Gołębiewski$^{a}$ and Maciej Krawczyk$^{a}$\chapaff{$^{a}$Institute of Spintronics and Quantum Information, Faculty of Physics, Adam Mickiewicz University, Uniwersytetu Poznańskiego 2, 61-614 Poznań, Poland\\
mateusz.golebiewski@amu.edu.pl}}

\section{Introduction}\label{sec1.1}

The exploration of spin waves (SWs) and their interactions within magnetic materials covers the interdisciplinary fields of magnonics and spintronics~\cite{Kruglyak2010Magnonics,Chumak2019MagnonSpintronics,Dieny2020}. This convergence has opened avenues for advanced research, particularly in harnessing SWs for signal transport avoiding Joule-Lenz heat emission~\cite{Serga2010YIGMagnonics, Yan2013MagnonWires, Barker2016ThermalGarnet}. The spectrum of SW wavelengths, ranging from micrometers to tens of nanometers, and corresponding to frequencies from a few to hundreds of GHz~\cite{Schneider2008, Maendl2017}, also promises advances in microwave technologies with a versatile platform for manipulating their dispersion and group velocities~\cite{Garcia-Sanchez2015NarrowWalls, Wagner2016MagneticNanochannels, Duerr2012EnhancedWaveguide, Lan2015Spin-WaveDiode, Krawczyk2014Review}. This versatility, combined with high energy efficiency without sacrificing processing speed~\cite{Chumak2015MagnonSpintronics, Kruglyak2010Magnonics, Serga2010YIGMagnonics, Lenk2011TheMagnonics,Mahmoud2020IntroductionComputing} positions SWs as a compelling alternative to traditional electric currents, especially in computing and microwave systems.

The design of advanced magnonic systems integrates multiple factors, such as geometry, topology, and material properties. Together with the magnitude and direction of the external magnetic field, these factors influence key properties of SWs, including supported SW frequencies, their anisotropy, and ferromagnetic response. This integrated approach enables precise control of magnon dynamics for targeted applications. Magnonic crystals (MCs) represent an innovative intersection of magnetic film technology and periodic patterning, serving as a platform for exploring wave dynamics in both two-dimensional (2D) and three-dimensional (3D) domains~\cite{Krawczyk2014Review,Chumak_2017}. In the case of 2D MCs, the characteristic in-plane dimensions usually span several hundred nanometers, while the magnetic films maintain thicknesses on the order of tens of nanometers~\cite{Rychly2015,Mamica2Dcrys}. In these ferromagnetic materials, the dynamics of SWs are primarily governed by isotropic exchange and anisotropic magnetostatic interactions~\cite{Krawczyk_2013}, making the physics of the system more complex than that of their photonic (light-based)~\cite{BUTT2021} and phononic (sound or elastic wave-based)~\cite{Vasileiadis2021} counterparts. Similar to other periodic composites, MCs allow for customization of the dispersion relation through structural and material composition adjustments, allowing for precise control over the velocity, direction, and magnonic gap width of propagating SWs~\cite{Krawczyk2014Review}. However, magnetism provides a broad mechanism for manipulating the dynamics of SWs that is not present in other types of artificial crystals and waves. 
A notable aspect of MCs is the way the SWs are affected by the geometry of the system through the internal magnetic fields. This is called the demagnetization field and can be tuned by the orientation of an external magnetic field, often resulting in a reduction of symmetry compared to the original structural design~\cite{Neusser2011,Tacchi2015,Gross2021,Li2022,MAMICA2023}. The other possibility, unique to MC, is the modulation of the SW properties by the magnetization texture, with or without nanostructuralization. For example, the magnetization domain structure in the form of periodic stripes can have a periodicity up to 100 nm, forming a periodic potential for the propagation of SWs, which can be considered as a fully reprogrammable MC~\cite{Banerjee2017,GRUSZECKI2019,YU20211,Petti_2022,Szulc2022}. In 1D and 2D, the MC can be also formed by the chain~ \cite{Ma2015,Mruczkiewicz2016,szulc2024} or the array~\cite{Wang2020,Tang2023} of skyrmions, due to the small skyrmion size, they are also stabilized with the period in deep nanoscale.

The MCs discussed above are predominantly planar structures that are uniform along the thickness, and due to the small thickness, the magnetization is also largely homogeneous. If the geometrical inhomogeneity or the magnetization texture becomes inhomogeneous along the thickness, we can call the system 3D~\cite{Beginin2019,Gubbiotti2021,Sadovnikov2022,Girardi2024}. Strictly speaking, however, 3D MCs are characterized by a three-dimensional periodic distribution of two materials consisting of inclusion and matrix. Enriched by the additional dimension, the properties of these systems were the focus of theoretical studies, including the exploration of magnonic bandgap phenomena~\cite{Krawczyk2006,Mamica2012,Romero2012}. They have demonstrated critical aspects such as the threshold of magnetic contrast required between the constituent materials to open a band gap and its dependence on the crystallographic structure of the crystal~\cite{Krawczyk2010}. In addition, it has been shown that the selection of an appropriate MC structure and filling fraction can lead to the design of anisotropic and strongly wavevector-dependent effective damping~\cite{Romero2012}, demonstrating the potential of 3D MCs to advance the field of magnonics and spintronics through nuanced control of wave propagation mechanisms in 3D space~\cite{Gubbiotti2019Three-DimensionalMagnonics}.

In addition to propagating waves that create bands in an MC and occupy the whole magnetic volume, an important aspect of the nanoscale magnonic systems is also the phenomenon of SW localization inside the 3D nanostructure or at the surfaces~\cite{Serha_DE, Liu2020}. A well-described type of SW localization already occurs on the surface of thin films in the Damon-Eshbach (DE) configuration~\cite{Damon1961,Hurben1995}, i.e., when the propagation of SWs is perpendicular to an applied external magnetic field that lies in the plane of the film and saturate the sample. The interplay between the dynamic stray fields from the surface and volume magnetic charges formed by oscillating magnetization in the film geometry leads to an asymmetry in the internal magnetic field distribution, creating a gradient in the effective magnetic field across the thickness (the non-zero wavenumber $k$ is required for this effect). As a result, the SWs have a higher intensity near one of the surfaces of the film compared to the inner part and opposite surface. The intensity of these surface-localized modes decays exponentially from the surface of the film toward its center, with the decay length depending on the SW wavelength, i.e., being proportional to $k$. The side of localization changes with the reversal of the direction of the wavevector (or magnetic field orientation). This type of SW localization offers several advantages for magnonic applications, e.g., increasing the efficiency of SW excitation with inherent nonreciprocity and detection at the film surfaces~\cite{Schneider2008,Jamali2013}, which is particularly beneficial for devices that rely on surface-based SW manipulation~\cite{Inoue2011,Bessonov2015}. In addition, confining the SWs to the surfaces reduces volumetric scattering and damping, potentially leading to lower energy loss and improved propagation characteristics~\cite{Yamamoto2019}. Finally, the DE configuration enables the design of devices with directional propagation properties, i.e., nonreciprocal, which can be exploited in the development of directional magnonic waveguides and logic elements.

SW localization can also be caused by static demagnetization fields. A notable manifestation of this is the emergence of edge modes~\cite{Jorzick2002, Park2002, Bayer2006, Bailleul2003, Kruglyak2006}, where SWs are confined or propagated~\cite{Gruszecki2021,Gruszecki2022} along the edges of a system, to which static magnetization is perpendicular. The reduction of the internal magnetic field at the edges of the ferromagnetic element can be modified by the magnetization orientation and the shape of the edge, allowing to increase or decrease the localization of low-frequency SWs~\cite{golebiewskiCresc}. This makes the edge SWs sensitive to the properties at the very edge of ferromagnets, which allows us to use them for sensing the magnetic properties~\cite{1Maranville2006,Maranville2007}. 

Another interesting type of localization is the topologically protected propagation of SWs along customized paths, facilitated by the MC edges or interfaces between two MCs with different topologies~\cite{Shindou2013,Shindou2013b,Li2018,McClarty2022,Zhuo2023}. These edges or interfaces act as ideal waveguides, guiding the SWs in one direction and being immune to backscattering, resulting in high coherence even along sharply curved paths~\cite{Wang2018,Feilhauer2023}. Such topological phases are characterized by so-called topological invariants~\cite{Kane2005, Kane2005-2, Konig2007}, such as Chern numbers, which describe the topological connection between bulk and boundary dynamics~\cite{Bansil2016}. Recently, an extension to higher-order topological states has become cutting-edge research~\cite{Schindler2018}, realized especially with the use of the artificial crystal in photonics~\cite{ElHassan2019} or acoustics~\cite{Ni2019}, and only initiated in magnonics showing corner SWs in 2D antiskyrmion crystal~\cite{Hirosawa2020}, and remains completely unexplored in 3D magnonic systems.

The study of 3D MCs holds great promise not only for SW localization phenomena. By tuning the geometry and lattice periods of their unit cells, these structures introduce an additional dimension for novel interactions, including various topological and geometric effects, and emergent material properties~\cite{Gubbiotti2019Three-DimensionalMagnonics, Fischer2020LaunchingNanostructures, Makarov2022NewNanoarchitectures, Cheenikundil2022, Fernandez-Pacheco2017Three-dimensionalNanomagnetism}. Recent advancements in techniques such as X-ray vector nanotomography~\cite{Donnelly2017}, magnetic laminography~\cite{Donnelly2022ComplexNanostructures}, two-photon lithography~\cite{Hunt2020HarnessingNanoscale, vandenBerg2023CombiningNanostructures}, focused electron beam deposition~\cite{Skoric2020}, and block copolymer templating~\cite{Llandro2020VisualizingNetworks} have enabled the fabrication of complex 3D magnetic systems at the nanoscale, providing the potential for rapid development of the field. Intriguing effects have already been observed in nanorods arranged in diamond-bond networks~\cite{May2019, Stenning2023}, facilitating the analysis of near-degenerate states and laying the groundwork for reconfigurable magnonic devices. Recently, gyroids~\cite{Luzzati1967, Schoen1970InfiniteSelf-intersections, Han2018}, characterized by their chiral triple junctions and fully interconnected 3D network, have been fabricated from ferromagnetic metals~\cite{Llandro2020VisualizingNetworks}. Due to the interplay of geometric, chiral and potential topological properties in the nanoscale, the ferromagnetic gyroids can be regarded as 3D MCs possessing all the above-mentioned properties, making them an ideal structure for studying SW dynamics~\cite{golebiewski2024gyr}. In this chapter, we summarize the current state of research on ferromagnetic gyroids and present our preliminary results on SW dynamics within these structures.

\section{Geometric properties of gyroidal networks}

Since their discovery in 1970~\cite{Schoen1970InfiniteSelf-intersections}, gyroids have been of interest to research in mathematics, material engineering, and photonics. These structures, belonging to the $\textit{I}4_1 32$ space group (No.~214)~\cite{Lambert1996TriplyMorphologies}, owe their unique morphology to the cooperative formation at inorganic-organic interfaces, as evidenced in surfactant-silicate mesostructures~\cite{Monnier1993Science}. This architecture has inspired a variety of studies in photonics, where gyroids have been employed as chiral beamsplitters, nonlinear optical metamaterials, and photonic crystals~\cite{Turner2013, Vignolini2012ASelf-Assembly, Dolan2015, Michielsen2008GyroidCrystals}. Recent advancements in fabrication techniques (see Introduction~\ref{sec1.1}) have also facilitated the creation of artificial systems inspired by gyroid geometry~\cite{Yan2012EvaluationsMelting, Yanez2016, Turner2013}.

\begin{figure}[htp]
\includegraphics[width=\linewidth]{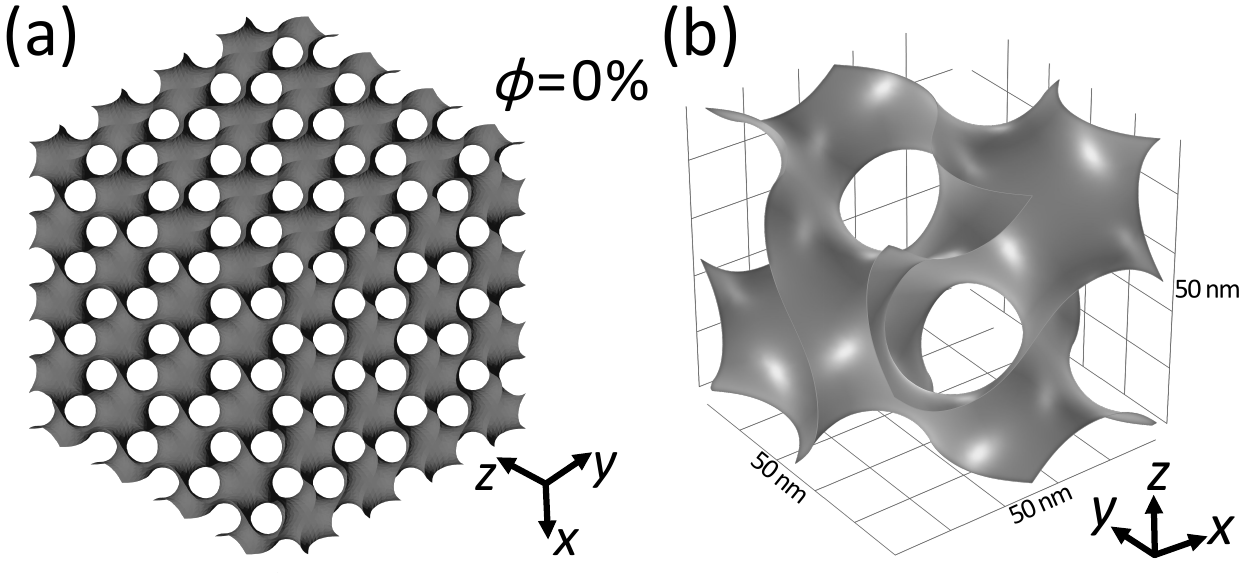}
\caption{The gyroid surface model, illustrating the arrangement along the [111] crystallographic direction through an orthographic projection (a), revealing the hexagonal patterning inherent to the interconnected channels of the system. Panel (b) shows the unit cell in a perspective projection, providing a dynamic view of its geometric configuration.
\label{Fig:0proc}}
\end{figure}

As detailed in Refs.~\cite{Schoen1970InfiniteSelf-intersections,Schoen2012reflections, Rosi2020hal}, the gyroid is characterized as a unique triple periodic minimal surface (see Fig.~\ref{Fig:0proc}). Its most notable property is its zero mean curvature, meaning that each point on the surface functions as a saddle point with equal and opposite principal curvatures~\cite{Dacorogna2014calculus}. This surface extends periodically along three orthogonal vectors and exhibits inherent chirality, lacking any symmetry plane or center, yet retaining rotational symmetry elements~\cite{Wohlgemuth2001triply}.
Due to its triple periodic nature, the gyroid is conceptualized as a crystalline structure~\cite{Chen2013acoustic}, adhering to the body-centered cubic (bcc) Bravais lattice and associated with the point group $O$, or described as $I4_{1}32$ in Hermann-Maugin notation. This cubic group, being purely rotational, emphasizes the chiral nature of the gyroid, and the prefix $I$ indicates a body-centered arrangement, suggesting a non-primitive, body-centered conventional unit cell.

Historically, the gyroidal surface was first described using the conjugate surface construction method~\cite{Karcher1989TheCompanions}, with its embedding later confirmed theoretically in Ref.~\cite{Groe-Brauckmann1996TheCompanions}. Subsequent studies dealt with the volume fractions of gyroids, especially those with constant mean curvature~\cite{Groe-Brauckmann1997GyroidsCurvature}. In a broader scientific context, the gyroid is recognized as Laves' graph of girth ten~\cite{Sunada2008CrystalsCreating} and the $K_4$ crystal~\cite{Hyde2008AMathematics, Mizuno2019RecentResearch}, known for their efficient space-filling properties and high symmetry. The connection to Laves' graph comes from the gyroid's intricate network of vertices and edges, which mimics the uniformity and connectivity of the graph. Similarly, the local arrangement of atoms in materials that form gyroid structures may be analogous to the tetrahedral coordination of the $K_4$ crystal.

Its complex design features cubic unit cells interconnected by elliptical cross-section nanorods, as detailed in Ref.~\cite{Dolan2015}. With a volume filling fraction of $\phi=0\%$, shown in Fig.~\ref{Fig:0proc}, the gyroid surface divides space into two distinct labyrinths, intersecting at 70.5~deg. Their mathematical representation is given by the trigonometric equation:
\begin{equation} \label{gyr_tryg}
\begin{split}
    \sin{(2\pi x/L)}\cos{(2\pi y/L)}&+\\ 
    \sin{(2\pi y/L)}\cos{(2\pi z/L)}&+\\ 
    \sin{(2\pi z/L)}\cos{(2\pi x/L)}&\le(101.5-2\phi)/68.1,
\end{split}
\end{equation}
where $L$ represents the unit cell dimension of the gyroid. For a more detailed derivation see e.g.~\cite{Tselikas1996}.

\subsection{Block copolymer self-assembly and fabrication techniques \label{sec:assembly_fabr}}

\begin{figure}[htp]
\includegraphics[width=\linewidth]{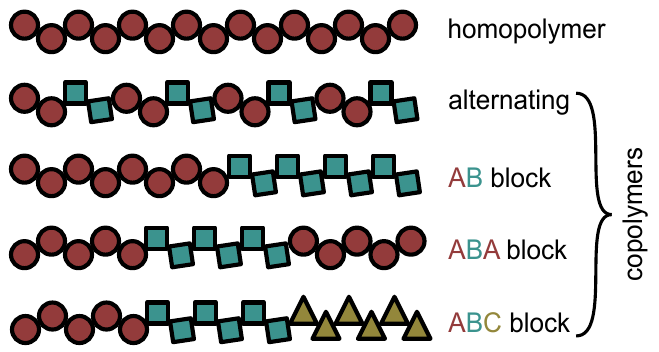}
\caption{Schematic illustration of various polymer types. Block copolymers consist of two or more homopolymer segments (A) connected by covalent bonds. Diblock copolymers contain two distinct segments (AB), while triblock copolymers have three (ABC). A block is technically defined as a segment of a macromolecule, made up of numerous repeating units, with at least one characteristic absent in adjacent segments.
\label{Fig:polymers}}
\end{figure}

\begin{figure}[htp]
\includegraphics[width=\linewidth]{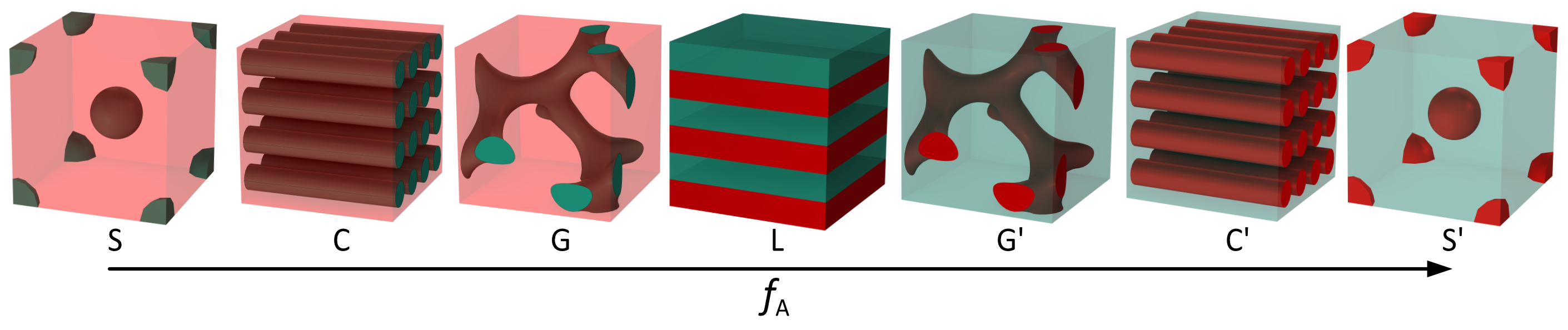}
\caption{Equilibrium morphologies of AB diblock copolymers in a bulk, varying as a function of the volume fraction $f_\text{A}$ of the blocks and the interaction parameter $\chi$. The morphologies are categorized as follows S/S' -- body-centered cubic spheres, C/C' -- hexagonally packed cylinders, G/G' -- bicontinuous gyroids, and L -- lamellae.
\label{Fig:poly_blocks}}
\end{figure}

Gyroidal ferromagnetic structures can be fabricated through the self-assembly of block copolymers, which naturally form intricate, periodic morphologies at the nanoscale. Block copolymers are composed of two or more different polymer segments or blocks, covalently bonded in a linear sequence, as presented in Fig.~\ref{Fig:polymers}. These blocks are derived from different types of monomers, small repeating molecular units, allowing for a wide range of chemical compositions within a single polymer chain~\cite{Feng2017}. The inherent property of block copolymers is their ability to spontaneously organize into well-defined nano- to microscale structures when subjected to certain conditions, such as changes in temperature. This self-assembly results from the incompatibility between the different blocks, leading to the formation of ordered morphologies such as spheres, cylinders, lamellae and complex 3D structures within a continuous matrix -- Fig.~\ref{Fig:poly_blocks}. The architecture of the block copolymer (e.g., diblock, triblock, multiblock) and the volume fraction of each block are critical factors that influence the resulting morphology.

Due to their customizable chemical compositions and the ability to form a wide array of structures, block copolymers have found applications across numerous fields. They can be tailored for use in nanoscale templates for electronic devices, photonic crystals, drug delivery systems, and materials with specific mechanical, optical, or conductive properties. The versatility and tunability of block copolymers, due to the variety of monomers from which they can be made and their unique self-assembly properties, make them a powerful tool in materials science and nanotechnology. For a comprehensive overview, refer to \cite{Dolan2015, HYDE1997141, HYDE199787, HYDE19971, SEGALMAN2005191, Cheng2006}.

A simple example of covalently bonded homopolymers are the linear AB diblock copolymers, representing the essential behavior of more complex systems like linear triblock terpolymers. The morphology of diblock copolymers in thermodynamic equilibrium is influenced by three key parameters: $N$ (total degree of polymerization, i.e., the number of monomer units in a polymer chain), $f_\text{A}$ or $f_\text{B}$ (volume fractions of each homopolymer), and $\chi$ (the Flory-Huggins interaction parameter, indicating monomer repulsion). In block copolymers, a higher $\chi$ value signifies greater incompatibility between the blocks, driving them to segregate and self-assemble into complex structures to minimize the free energy of the system, e.g., gyroids (Fig.~\ref{Fig:poly_blocks} -- G/G'). The interplay between $\chi$, which promotes phase separation, and $N$, which influences the entropy and enthalpy of the system, guides the self-assembly process.

As the block copolymers self-organize, the gyroid structure emerges as a result of the balance between the repulsive interactions (dictated by $\chi$) and the chain length of the polymers (defined by $N$). This balance leads to the formation of continuous, triple periodic minimal surfaces, which are characteristic of the gyroid structure. This architecture, with interconnected labyrinths of both materials, is thus a direct consequence of the molecular properties of the block copolymers and their drive to achieve a thermodynamically favorable configuration. Larger gyroid structures, scaling from millimeters to sub-millimeters, are typically fabricated using top-down methods~\cite{Yan2012EvaluationsMelting, Yanez2016, Turner2013, Gan2016}. Conversely, the creation of nanostructures with unit cells below 100~nm is predominantly achieved through bottom-up techniques~\cite{Armatas2006, Kresge1992OrderedMechanism}. The process involves microphase separation in copolymers, resulting in gyroid networks of a minority polymer block within a majority block matrix~\cite{Ross2014, She2013}.

\begin{figure}[htp]
\includegraphics[width=\linewidth]{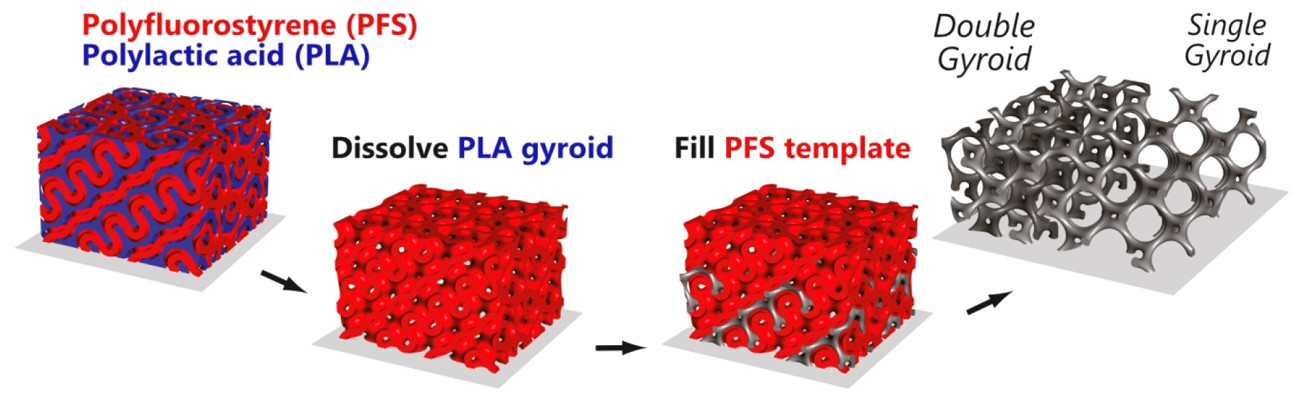}
\caption{Fabrication and structural characterization of gyroid nanostructures. This scheme shows the process for creating free-standing metallic gyroid nanostructures, involving the thermal annealing of a block copolymer template, followed by the selective removal of the minority block and subsequent electrodeposition. Reprinted (adapted) with permission from \textit{Nano Lett. 2020, 20, 5, 3642-3650}. Copyright 2024 American Chemical Society.
\label{Fig:fabrication}}
\end{figure}

The formation of complex structures using block copolymers as templates involves a multi-step process that exploits the self-assembly properties of block copolymers and the technique of electrodeposition to introduce ferromagnetic materials, such as permalloy (Py) or nickel (Ni), into the predetermined patterns. One of the polymer blocks is selectively removed to create a porous template that retains the architecture. This can be achieved by various methods, including chemical etching or UV degradation, depending on the nature of the block copolymer.
With the porous template in place, electrodeposition is used to fill the voids with a ferromagnetic material, as illustrated in Fig.~\ref{Fig:fabrication}. In this instance, a diblock copolymer of poly(4-fluorostyrene) (PFS) and 38\% poly(lactic acid) (PLA) is employed, with PFS and PLA serving as the majority and minority blocks, respectively~\cite{Scherer2014, Llandro2020VisualizingNetworks}. During electrodeposition, the template is immersed in an electrolytic solution containing ions of the ferromagnetic material. When a voltage is applied, the ions are reduced and deposited onto the conductive areas of the template, gradually filling the structure. After the electrodeposition process is complete, the remaining block copolymer template is removed, often by solvent washing or thermal decomposition, leaving a freestanding structure composed entirely of the ferromagnetic material. The end product is a 3D structure of ferromagnetic material that replicates the geometry of the original template.

In Fig.~\ref{Fig:fabrication}, the fabrication of magnetic (based on Ni) gyroid nanostructures from Ref.~\cite{Llandro2020VisualizingNetworks}, featuring unit cell dimensions smaller than 50~nm, is demonstrated. These nanostructures are realized through electrodeposition into block copolymer templates, yielding $\text{Ni}_{75}\text{Fe}_{25}$ gyroids. 
The self-assembly of 3D nanostructures, especially when integrated with inorganic components~\cite{Simon2001BlockCopolymer}, allows unprecedented morphological control at the nanoscale, opening new avenues for magnetism and magnonics. The SW interactions within gyroids are complex and involve, among others, shape anisotropy, inhomogeneous demagnetization fields, curved surfaces, and chirality. Our research aims to show the complexity of these interactions, provide an understanding of the magnetic behavior in gyroids, and lay the framework for future explorations and use of 3D ferromagnetic magnonics.

\section{Gyroid structures in magnetism}

Short SWs, with their nanoscale wavelengths and microwave frequencies, enable high-density, high-frequency applications of miniaturized magnonic devices~\cite{Chumak2015MagnonSpintronics}. The concepts of SWs in connection with the chirality inherent in the magnetization dynamics, as discussed above and reviewed in Ref. ~\cite{Kruglyak2021}, on the one hand, and introduced by the Dzyaloshinskii-Moriya interaction (DMI)~\cite{Tacchi2023} or geometric curvature~\cite{Otalora2017,Sheka2022}, on the other hand, may play an important role in the further development of information processing technologies based on magnonics~\cite{Chen2020,Chen_2022}. Chirality, which refers to the preference of spin structures to twist in a particular direction, facilitates also directional SW propagation, opening the door to non-reciprocal devices. DMI, which exists in systems with broken inversion symmetry~\cite{Gan2016}, stabilizes chiral magnetic textures, improving control over spin structure and enabling the design of robust, energy-efficient memory devices, the structures which can also be exploited for magnonics~\cite{Garcia2014}. Similar properties are expected to exist in curvilinear ferromagnetic systems, since curvature, when introduced on a length comparable to the exchange length, introduces the anisotropic exchange in the same form as DMI~\cite{Sheka2022}. 
Therefore, the complex architecture of gyroidal structures, with the 3D gyroid elementary cell characterized by curvature and chirality (see Figs.~\ref{Fig:0proc} and \ref{Fig:gyr_unit_cell}), provides a unique, three-dimensional platform for exploiting all of these phenomena, making it a good candidate for further research, and promising applications.
Nonetheless, the study of magnetization dynamics within periodic 3D nanostructures is still in its early stages. While extensive research exists on SWs in uniform, 1D and 2D structures, investigations into collective SW dynamics in 3D artificial systems~\cite{May2021MagneticSpin-ice, Grundler2023,Gubbiotti2019Three-DimensionalMagnonics,Fernandez-Pacheco2017Three-dimensionalNanomagnetism}, particularly gyroids~\cite{golebiewski2024gyr}, remain limited. We attempt to summarize existing and explore new research directions on the collective dynamics of SWs in gyroidal nanostructures.

As described above, the inherent chirality and curvature of nanoscale magnetic gyroids offer a promising pathway for controlling non-collinear spin textures. Recent studies have underscored this potential~\cite{Lich2023FormationNanostructures}, and the experimental visualization of magnetic structures in $\text{Ni}_{75}\text{Fe}_{25}$ gyroid networks has provided further evidence~\cite{Llandro2020VisualizingNetworks}.
However, the DMI and curvature-induced anisotropy, which add another layer of complexity to research in 3D ferromagnetic structures~\cite{hertel_curvature-induced_2013, gaididei_curvature_2014}, have not yet been explored in gyroidal structures. Similarly, novel physical effects observed in curved magnetic wires and films~\cite{Sheka2021AMagnetism}, and the integration of chiral and topological properties~\cite{Shindou2013, McClarty2022}, have not been tested in gyroids. Furthermore, the potential of gyroids extends further, with the numerous energetically equivalent stable states, suggesting suitability for the realization of artificial spin-ice systems in 3D. 
Indeed, gyroidal magnetic nanostructures, particularly those with at least one dimension in the exchange length scale, have emerged as key to the realization of in-volume system phenomena such as monopole-type excitations~\cite{Wannier1950,diep2013frustrated,
lacroix2011introduction}, which were previously observed only in two-dimensional (2D) systems~\cite{Sandra2020, Nisoli2017}. 

Taking the magnonic perspective, we use micromagnetic simulations to reveal the collective dynamics of SWs in gyroids. This approach is consistent with the findings in Ref.~\cite{Demidov2010Oscillators}, where the importance of direct observation and mapping of SWs, especially in nano-oscillators, is highlighted. An interesting aspect of our investigations is to study the ferromagnetic resonance and SW dispersion within gyroids. The frequency-wavevector relationship of SWs is of particular interest in complex 3D geometries, where the multidimensional nature introduces additional complexities in SW propagation, potentially leading to tunable, magnetic field-directed band gaps, mode crossings/hybridizations, and unconventional group velocities. By analyzing these properties, we aim to unveil the potential of gyroids as magnonic metamaterials and efficient SW carriers.

\subsection{Static magnetization configuration}

\begin{figure}[htp]
\includegraphics[width=\linewidth]{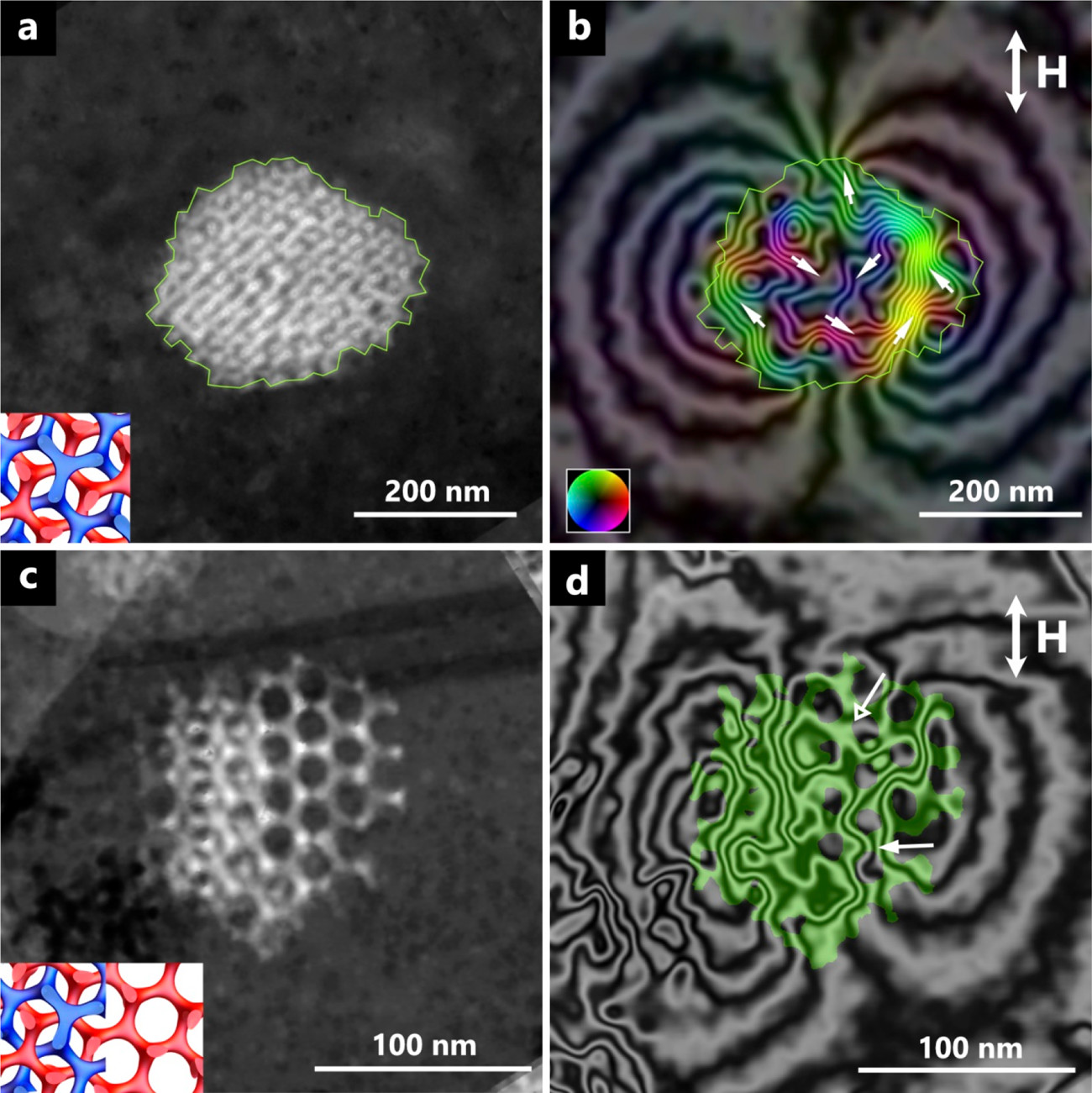}
\caption{Magnetographic analysis of Py double and single gyroid structures with unit cell size $L=42$~nm: insets illustrate simulated models of double-gyroids and hybrid double/single-gyroid structures, with each gyroid network marked in red and blue. Panels (a) and (b) display the phase contribution of the mean inner potential and the map of magnetic induction for a double-gyroids. Bidirectional arrows indicate the saturation magnetic field ($\textbf{H}$) direction. Stray magnetic fields around the structure are traced by contour lines, which within the struts is non-uniform, as indicated by unidirectional arrows representing varied field directions. The structure's outline, derived from the phase shift's mean inner potential contribution, is shown in (b). In panels (c) and (d), the phase contribution of the mean inner potential and the corresponding magnetic flux contour map are depicted for a sample containing both double-gyroid and single-gyroid regions. In (d), the gyroid particle is masked in green and superimposed with contour lines to illustrate the relationship between field lines and gyroid struts. Points where flux contours intersect gyroid struts or loop around vertices are marked with closed and open unidirectional arrows, respectively. The contour interval for both (b) and (d) is set at $2\pi/64$ radians. Reprinted (adapted) with permission from \textit{Nano Lett. 2020, 20, 5, 3642-3650}. Copyright 2024 American Chemical Society.\label{Fig:gyr_static}}
\end{figure}

The study of static magnetization in 3D gyroidal ferromagnetic nanostructures apart from learning the specifics of these structures, also can advance our understanding of complex magnetic systems in general, and open up new technological possibilities, as they have already been recognized for their intricate, frustrated remanent states~\cite{Llandro2020VisualizingNetworks}. These states are inherently ferromagnetic, but do not conform to a unique equilibrium configuration. The detailed imaging and mapping of gyroids, shown in Fig.~\ref{Fig:gyr_static}, provide a deeper insight into their magnetic structure, highlighting the complex interplay between the magnetic fields and the intricate single and double gyroid architecture. Figure~\ref{Fig:gyr_static}(b) shows the magnetic induction map of the double-gyroid nanostructure. The external stray field surrounding the gyroid exhibits dipole-like behavior aligned with the saturating field (\textbf{H}). Within the structure, multiple flux-closed loops are evident, some of which enclose only a single unit cell. However, identifying the chirality of these loops is challenging due to the integration of the magnetic induction map across the thickness of several unit cells and the entangled nature of the double-gyroid networks.

In the single-gyroid region [Fig.~\ref{Fig:gyr_static}(d)], the correlation between the structure and the magnetization of the ferromagnetic gyroid becomes more pronounced. The external stray field remains dipole-like, but the remanent state of the single gyroid is more complex than a simple network of flux-closed loops. The contours are generally aligned with the structure, but in some regions (marked with the arrows) the flux lines cut directly across the struts, implying that the magnetization is transverse to the strut axis. In other regions, the flux contours encircle specific vertices, possibly indicating that the magnetization of the three constituent struts converge or diverge from the vertex center.

In Ref.~\cite{Lich2023FormationNanostructures}, the authors have presented an insightful analysis through phase-field simulation studies, focusing on the formation and switching behavior of magnetization textures within Py gyroid nanostructures with a periodicity of 50~nm. The study uncovers the coexistence of left-handed (LH) and right-handed (RH) helices in these structures, a feature particularly pronounced in gyroids with small solid volume fractions. Moreover, it has been demonstrated that at small $\phi$, the magnetization textures conform to the ice rule. However, as $\phi$ increases, this rule is disrupted, leading to magnetic field texture frustration, a phenomenon that becomes prominent at a critical volume fraction of about 30\%. Furthermore, the findings indicate an increase in the coercive field as the volume fraction rises from 10 to 50\%, followed by a reversal of this trend with further increases in $\phi$. This behavior is attributed to the balance between LH and RH magnetic helices, with global switching events occurring near $\phi=50$\%, while local switching is observed in other volume fractions. They summarize the relationship between solid volume fraction and static magnetization distribution in gyroids.

Exploring the realm of static magnetism within complex 3D structures~\cite{hsueh2011_NanoporousGyroid} opens up promising avenues of research, particularly in the creation and control of topological spin textures such as skyrmions, torons~\cite{Li2022} and hopfions~\cite{Tai2018} and others~\cite{Wu2022}. The curvature and topology inherent in gyroids provide a novel platform for studying how these structures can stabilize such textures. Additional research focusing on static magnetization configurations under external fields could deepen our understanding of domain distribution and orientation. Also, the process of magnetization reversal in gyroids, influenced by geometry and thermal fluctuations, could provide valuable insights into material optimization for precise magnetic switching. Broadening the scope to include hybrid structures by integrating materials such as superconductors or topological insulators with gyroidal ferromagnets opens up other research opportunities.

Studying the magnetization texture and dynamics at the level of individual nanostructures, unit cells, vertices and struts, is essential for a comprehensive understanding of their magnetic behavior. Current research suggests the possibility of multiple equivalent magnetization configurations within the 3D gyroid network, which is particularly promising for applications such as reservoir computing~\cite{TANAKA2019100}, where systems benefit from high interconnectivity and a variety of stable states. The ability to access and manipulate these different states could enable new paradigms in nanoscale computing by exploiting the properties of these complex magnetic 3D structures.

\subsection{Magnetization dynamics}
\begin{figure}[htp]
\includegraphics[width=\linewidth]{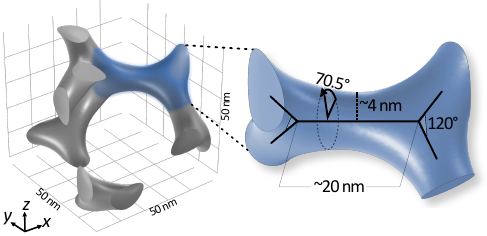}
\caption{Illustration of the cubic gyroid unit cell tailored for micromagnetic simulations, with dimensions $L=50$~nm and a volume fraction of $\phi=10\%$. An enlarged inset highlights the chiral linkage between two primary gyroid nodes, emphasizing key geometric features.
\label{Fig:gyr_unit_cell}}
\end{figure}

As already mentioned in the Introduction, the study of SW dynamics in 3D nanostructures is at a very early stage~\cite{Gubbiotti2019Three-DimensionalMagnonics,Cheenikundil2022,Li2023}, especially experimentally~\cite{Sahoo2018,Donnelly2020,Grundler2023,Girardi2024}. In the following subsections, we will summarize our research on Ni-based gyroids, including experimental ferromagnetic resonance (FMR) studies and numerical simulations. Here, the gyroid sample's unit cell is defined with a dimension of 50~nm and $\phi=10$\%, as shown in Fig.~\ref{Fig:gyr_unit_cell}. This specific dimensionality leads to strut diameters that align with crucial nanoscale measures, e.g., the exchange length (around 9~nm for Ni). This congruence highlights the profound impact of nanoscale features on the magnetic dynamics and properties we observe.

\subsubsection{Gyroidal crystallography vs. ferromagnetic response \label{sec:fmr}}

\begin{figure}[htp]
\includegraphics[width=\linewidth]{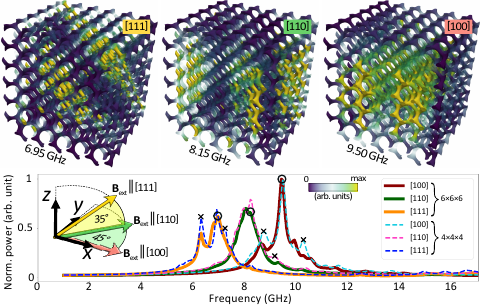}
\caption{Spectral analysis of SW modes in a $6\times6\times6$ gyroid structure. The lower part of the figure shows the frequency spectra with high-intensity volume modes highlighted by black circles. For comparison, spectra of more compact structures (dotted lines) are superimposed, showing a significant reduction in the intensity of edge modes (indicated by crosses) with increasing structure size. The color gradient in the mode visualization corresponds to the imaginary part of the magnetic susceptibility. Reprinted (adapted) with permission from \textit{ACS Appl. Mater. Interfaces 2024, 16, 17, 22177-22188} licensed under CC-BY 4.0. Copyright 2024 American Chemical Society.\label{fig:666_spectra}}
\end{figure}

Our recent work~\cite{golebiewski2024gyr} presents a synergistic approach combining micromagnetic simulations and FMR measurements, employed to examine the magnetic properties of three-dimensional gyroidal Ni nanostructures. This investigation reveals several distinctive characteristics of the gyroid network, particularly the pronounced influence of the static, external magnetic field orientation on the ferromagnetic response. This relationship is closely linked to the crystallographic alignment of the gyroid structure, as presented in Fig.~\ref{fig:666_spectra}. 

We utilize the finite element method (FEM) micromagnetic simulations in \textit{tetmag}, both in the time domain for determining static magnetization structures \cite{hertel_tetmag_2023}, and, using add-on algorithm, in the frequency domain for analyzing oscillatory, magnonic properties \cite{DAquino2023MicromagneticSystems} -- see more details in Appendix~\ref{Sec:sim}.
The following, Ni material parameters have been used: the saturation magnetization $M_\text{S}=480$~kA/m, the exchange stiffness $A_\text{ex}=13$~pJ/m, the Gilbert damping coefficient $\alpha=0.008$, and the $g$-factor of 2.14~\cite{Coey2010MagnetismMaterials,Singh1976CalculationNickel}.

We examine the impact of three specific field directions, aligned with the crystallographic axes [100], [110], and [111], on finite-sized gyroids, see Fig.~\ref{fig:666_spectra}. The findings from the resonance spectra simulations attribute the observed shifts in resonance signals in the rotating gyroid sample to crystallographic anisotropy. Moreover, several bulk SW modes, as identified in this work, exhibit a predominantly localization on the different regions across the perpendicular axis of the gyroid structure, on the facets perpendicular to the [110] and [100] magnetic field directions. This behavior likely stems from a combined effect of crystallographic influence and shape anisotropy impacting the localization of resonant modes across various regions, potentially influenced by the system's chirality. Furthermore, we identified some SW modes localized on the outer surfaces of the gyroid structure (see the modes marked with 'x' in the spectra shown in Fig.~\ref{fig:666_spectra}). This also indicates the important influence of the shape and outer surfaces of the 3D nanostructure on the SW dynamics. This observation prompts us to investigate the surface effects further. The preliminary results of our numerical studies of surface-localized SW modes are summarized in the next section.

\begin{figure}[htp]
\includegraphics[width=\linewidth]{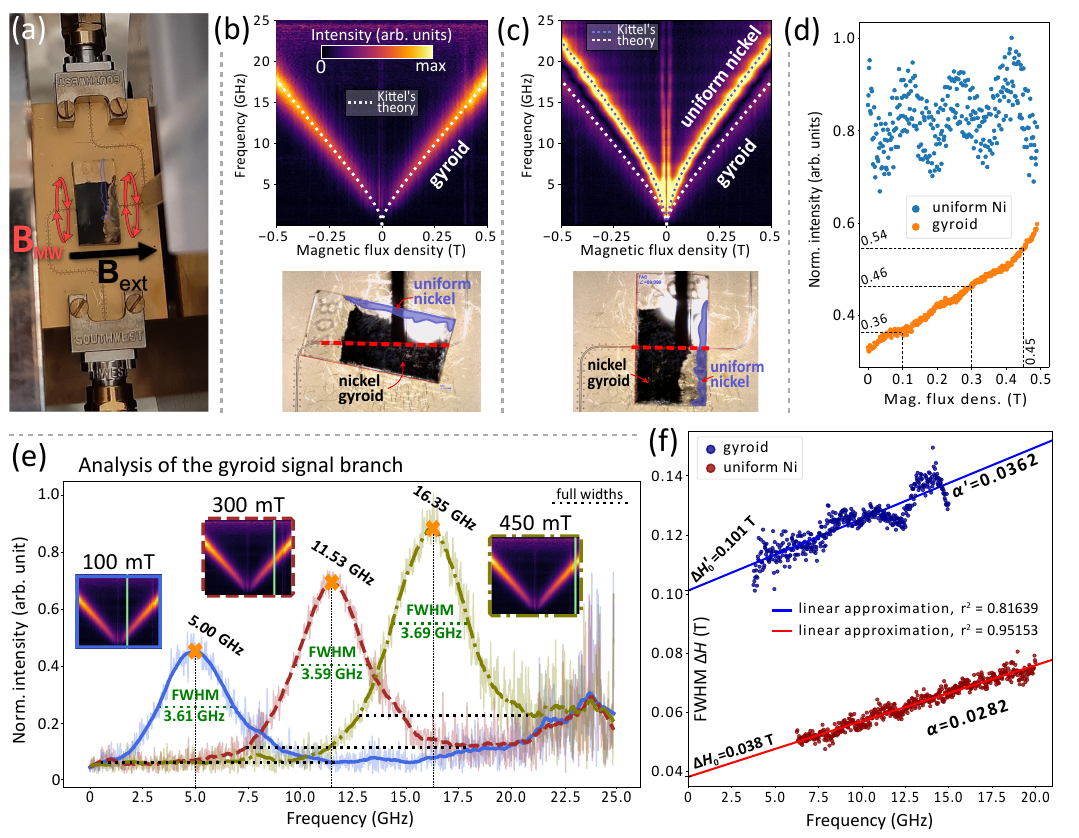}
\caption{A broadband FMR measurement on a Ni gyroid structure. The sample was repositioned relative to the CPW to study the effect of an additional homogeneous Ni layer. Two configurations were used to assess energy absorption from the microwave field $B_\text{MW}$, applied perpendicular to the static magnetic field (a). Dynamic magnetization amplitude as a function of static magnetic flux density and frequency for selected configurations are shown in (b) and (c). A strong signal from the gyroid layer appears when the CPW aligns directly below it (b), while a higher-frequency signal from the uniform Ni layer is observed when the CPW intersects its projected position (highlighted in purple) (c). Dotted lines represent theoretical fits using the Kittel formula (Eq.~\ref{eq:kittel}). For the uniform Ni, parameters from micromagnetic simulations were used; for the gyroid, calculated effective parameters ($M_\text{eff}=132$~kA/m, $g$-factor of 2.2) were applied to the formula. Plot (d) summarizes peak intensities of FMR signals (blue dots for uniform Ni, orange dots for gyroid) as a function of external magnetic field strength, with normalized values for 100~mT, 300~mT, and 45~mT. Plot (e) provides a cross-sectional analysis of FMR signals at different external magnetic fields: solid blue line for $B_\text{ext} = 100$~mT, dashed brown line for $B_\text{ext} = 300$~mT, and dash-dotted green line for $B_\text{ext} = 450$~mT. Horizontal dashed green lines indicate the FWHM for each section. Orange crosses mark peak maxima and corresponding frequencies. Insets show intensity plots from FMR measurements with green vertical lines marking specific section locations. Plot (f) shows the magnetic field FWHMs as a function of frequency for FMR signals of gyroid (purple dots) and uniform Ni (dark red dots). Linear regression, based on experimental data and Eq.~\ref{eq:deltaH}, estimates determination coefficient $\text{r}^2$, $\Delta H_0$ (from the abscissa), and damping values $\alpha$ (from the slope). Parameters for the gyroid are marked with a prime (${}^\prime$). Reprinted (adapted) with permission from \textit{ACS Appl. Mater. Interfaces 2024, 16, 17, 22177-22188} licensed under CC-BY 4.0. Copyright 2024 American Chemical Society.\label{fig:gyr_exp}}
\end{figure}

The \textit{tetmag} simulations were complemented by meticulous experimental measurements using broadband FMR, where a multi-domain (multiple, interconnected sub-parts with different crystallographic orientations) gyroid sample was positioned on a coplanar waveguide (CPW) line. This setup facilitated the analysis of distinctions and correlations between a solid, uniform Ni layer and the gyroid-structured segment of the sample. The Ni gyroid nanostructure analyzed was fabricated through the thermal annealing of a block copolymer template [detailed in Sec.~\ref{sec:assembly_fabr} (Fig.~\ref{Fig:fabrication}) and Ref.~\cite{Llandro2020VisualizingNetworks}], followed by selective dissolution of one of the gyroid-forming blocks and filling the resultant right-handed gyroid network with Ni by electrodeposition ($L=50$~nm, $\phi=10$\%). Performed measurements and provided numerical analysis allow us to conceptualize gyroid films as homogeneous materials or magnonic metamaterials, where the effective saturation magnetization is reduced in proportion to the gyroid's filling factor -- see Fig.~\ref{fig:gyr_exp}. We can see, that the gyroid signal exhibits a broader frequency full-width at half-maximum (FWHM) compared to that of homogeneous Ni. For instance, at a magnetic field of 450~mT, the FWHM is 3.69~GHz for the gyroid and 3.09~GHz for Ni. This broader signal is likely due to the multidomain nature of the sample. Specifically, the observed signal from the gyroid represents an average from multiple domains with varying crystallographic orientations located above the CPW. From the FMR measurements shown in Fig.~\ref{fig:gyr_exp}(e), we deduce an average FWHM difference of 1.3~GHz across various external magnetic field values. Micromagnetic simulations for a cube-shaped sample (Fig.~\ref{fig:666_spectra}) indicate a maximum peak separation of 2.95~GHz at 300~mT, specifically between the [100] and [111] crystallographic directions. These simulations evaluated the crystallographic orientations that represent the most divergent configurations of the gyroid lattice relative to the applied field, resulting in the largest possible separation of resonant frequencies. 
In field-swept FMR experiments conducted at a fixed frequency, the absorption line FWHM conforms to the equation $\mu_0\Delta H=4\pi\alpha f/\gamma$. This relationship holds when the magnetization vector is aligned with the applied magnetic field, either in-plane or perpendicular to it, leading to a linewidth that scales proportionally with frequency. The slope of this scaling is defined by the Gilbert damping parameter $\alpha$. In addition to this intrinsic contribution, empirical data suggest the presence of a frequency-independent term (see Eq.~\ref{eq:deltaH}), which contributes to the overall linewidth observed in the experiments. Experimental investigations have also enabled a linear regression analysis of FWHM across a spectrum of $B_\text{MW}$ frequencies, as depicted in Fig.~\ref{fig:gyr_exp}(f) for the designated orientations of the sample over CPW. This analytical approach is instrumental in ascertaining the damping values $\alpha$ and the inhomogeneous linewidth contribution $\Delta H_0$ for both the homogeneous Ni layer and the gyroidal structure, respectively. In the first case, the derived values are $\alpha=0.0282 \pm 2.54\%$ and $\Delta H_0=0.038 \pm 1.28\%$~T. Comparatively, the gyroidal structure exhibits $\alpha'=0.0362 \pm 6.49\%$ and $\Delta H'_0=0.101 \pm 1.05\%$~T. Larger $\alpha'$ for the gyroid structure is likely attributable to the scattering of SW modes within the nanowires, which are much thinner than the bulk Ni and exhibit complex noncollinear interconnections.

\subsubsection{Localization properties \label{sec:localization}}

\begin{figure}[htp]
\includegraphics[width=\linewidth]{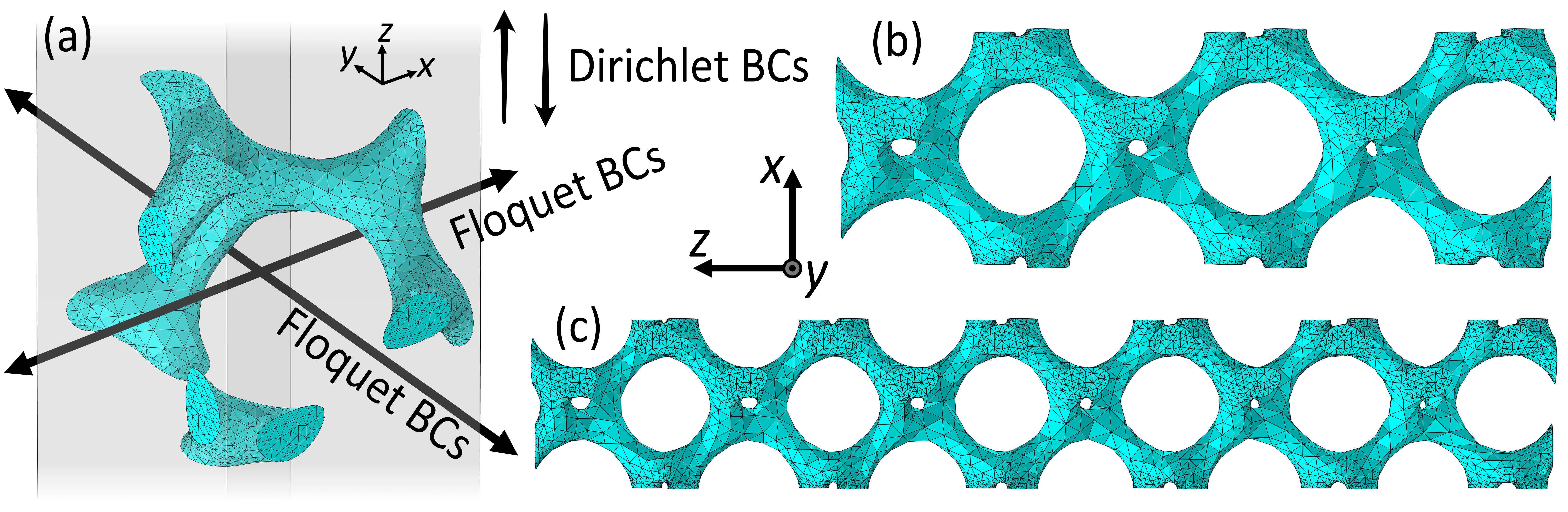}
\caption{Visualization of discretization grids for gyroid structures utilized in Comsol micromagnetic simulations. Arrows indicate the Bloch-Floquet boundary conditions (BCs) along the $x$ and $y$ axes, defining the structure's plane. In contrast, along the $z$-axis, perpendicular and at a significant distance from the plane, Dirichlet boundary conditions (BCs) are implemented. This setup is exemplified for a single-unit thick ($1\times1\times1$) gyroid layer in (a), and similarly for other models: $1\times1\times3$ in (b), and $1\times1\times6$ in (c). Note that the relative sizes of the structures are not to scale in this representation.
\label{Fig:gyr_sizes}}
\end{figure}

To study surface effects in gyroids, we simulate the gyroid in a thin film geometry in dependence on the thickness of the gyroid structure and the orientation of the external magnetic field. To avoid the influence of the lateral edges effects on the SW spectra we employ Bloch-Floquet boundary conditions at the unit cell boundaries. The simulations are performed with Comsol Multiphysics software. In this implementation, we assume full magnetization saturation and solve the Landau-Lifshitz-Gilbert equations in the linear approximation as an eigenproblem. Further details about this numerical approach can be found in Refs~~\cite{Mruczkiewicz2013StandingCrystals,Rychy2018SpinRegime} in 2D implementation, and in the Appendix~\ref{Sec:sim}. With that, we model an infinite gyroidal plane with varying thicknesses: 1, 3, and 6 unit cells, corresponding to dimensions of 50, 150, and 300~nm, respectively (Fig.~\ref{Fig:gyr_sizes}). Material parameters identical to those in the previous section were used. Throughout our simulations, the external magnetic field, $\textbf{B}_\text{ext}$, was maintained constant at 500~mT and oriented in the plane of the gyroidal plane along the crystallographic direction [100]. This field strength, corroborated by Ref.~\cite{Llandro2020VisualizingNetworks}, supports our assumption of complete saturation of the structure aligned with the external magnetic field's direction.

\begin{figure}[htp]
\includegraphics[width=\linewidth]{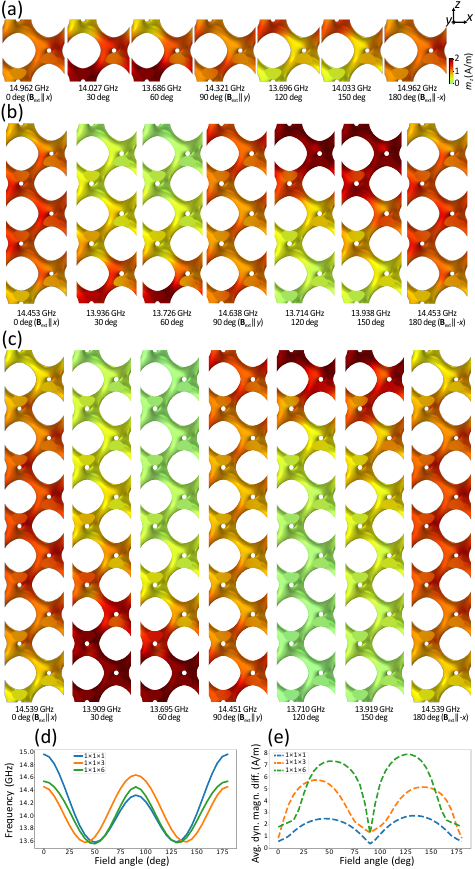}
\caption{Distribution of low frequency, the most intensive mode in (a) $1\times1\times1$, (b) $1\times1\times3$, and $1\times1\times6$ gyroids ($\phi =10$\%) and their localization as a function of magnetic field rotation ($B_\text{ext}=0.5$~T) with respect to the $x$-axis. The plots below illustrate the impact of the rotating field's angle relative to the structure on the eigenfrequency (d) and the variation between the minimum and maximum dynamic magnetization components within the unit cell (e).
\label{Fig:loc}}
\end{figure}

During the analysis of the micromagnetic simulation results, a surprising pattern emerged in the behavior of the lowest frequency mode, which exhibited the highest intensity, particularly under the influence of the rotating in-plane magnetic field. It is, an intriguing switch in the area of localization of SWs from the bottom [at 30 and 60 deg in Fig.~\ref{Fig:loc}(b) and (c)] to the top of the layer (at 120 and 150 deg). Between these angles, specifically at 0, 90, and 180 degrees, this mode exhibits a bulk character. This shift in mode amplitude concentration regions suggests a complex interplay between magnetic field orientation and intrinsic properties of the gyroid structure, involving dipolar and exchange interactions. The change of the amplitude localization is followed by the change of the resonance frequency [see Fig.~\ref{Fig:loc}(d)], the lowest (about 13.6~GHz at the field angle of 45-50 deg and 125-130 deg) occurs when the mode has a surface character and increases with the transition into the bulk region.

Preliminary analysis indicates that this localization phenomenon is more pronounced in structures with increased thickness, i.e., characterized by a higher number of unit cells per thickness, as can be seen by comprising SW amplitude distribution in Fig.~\ref{Fig:loc}(a)-(c). This has been quantified by plotting the difference between the maximum and minimum value of the dynamic magnetization component in the unit cell in Fig.~\ref{Fig:loc}(d). The largest value of this difference indicates stronger amplitude localization. Interestingly, the angle at which maximum localization occurs also shifts with changes in film thickness. In such cases, the crystallography exerts a greater influence on the SW amplitude distribution and indicates a possible influence of chirality.

The findings described are intriguing and require further investigation to unravel the complex interactions that govern these phenomena. Understanding the dynamics controlling the localization of SW modes in these structures has the potential to open up new applications in 3D magnonics, such as sensing or controlling dynamical coupling by rotating the magnetic field.

\subsubsection{Dispersion relation \label{sec:disp}}

\begin{figure}[htp]
\includegraphics[width=\linewidth]{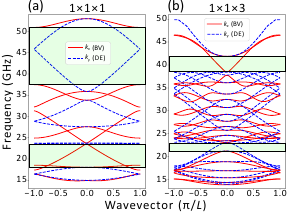}
\caption{The dispersion relations within the 1st Brillouin zone for a gyroid structure along the $x$-axis ([100] crystallographic direction), characterized by a 10\% filling factor and 1 (a), and 3 (b) unit cells per thickness. The data is color-coded to distinguish between different configurations: the BV configuration is represented in red, while the DE configuration is depicted in blue. These plots have been generated in the presence of an in-plane external magnetic field, $B_\text{ext}=0.5$~T, oriented along the $x$-axis. Wider band gaps specific to one of the configurations are highlighted using green rectangles.
\label{Fig:disp}}
\end{figure}

The surface character of the SW mode at $k=0$ has not been found in homogeneous thin ferromagnetic films so far. Thus, it is interesting to see how this and other modes in the gyroid structure evolve with the wavevector. The introduced Bloch-Floquet boundary conditions in Comsol MultiPhysics are very well suited for the calculation of the dispersion relation, just by parametrize the solution with the Bloch wavevector, $k$.
In reciprocal lattice space, the range $-\pi/L<k<\pi/L$ defines the 1st Brillouin zone along the main direction of the cubic lattice. The primary focus is on the 1st Brillouin zone, as the bands from the subsequent zones are folded back to the 1st zone, and the whole magnonic band structure has a periodicity with the period equal to the reciprocal lattice vector $2\pi/L$. Our research delves into the SW dispersion in a thin film based on the gyroid structure, defined in the previous section. It displays dispersion relations similar to those of traditional magnonic crystals but with specific features. As evidenced in Fig.~\ref{Fig:disp}, the direction of the magnetic field relative to SW propagation significantly influences the gyroid’s band structure, leading to the opening and closing of band gaps. This phenomenon is exemplified in our simulations of the $1\times1\times1$ [Fig.~\ref{Fig:disp}(a)] and $1\times1\times3$ [Fig.~\ref{Fig:disp}(b)] gyroid structures for two relative orientations of the in-plane magnetic field and wavevector, i.e., DE configuration in which both directions are orthogonal, and backward volume (BV) configuration in which these vectors are parallel. In the DE configuration, in $1\times1\times1$ structure, we observe a wide band gap from 18.1 to 23.6~GHz, limited by the flat band, and another bend gap just above, from about 23.8 to 27.9~GHz. On the other hand, in BV configuration, there are only two very narrow gaps at low frequencies and a wide band gap between 37.6~GHz and 51.2~GHz. The spectra change significantly when we increase the film thickness to 3 unit cells [Fig.~\ref{Fig:disp}(b)]. In this case, the band structure is much denser due to the decreasing frequency of the SW modes quantized along the thickness. The band gaps at low frequencies are now closed for DE, but it is opened for BV configuration, although it is rather small, only 1.8~GHz wide (around 22~GHz). In higher frequencies, the band gap in DE configuration opens (38.6-41.9~GHz). Interestingly, in the frequency range from 23 to 38.6~GHz, the DE bands show a pronounced slope with numerous crossings but no hybridizations, whereas in the BV configuration, extensive hybridizations result in a wavy band structure as a function of $k$, separated by distinct band gaps.

The preliminary analyses reveal the complexities of SW propagation in thin films based on gyroid structures, such as the emergence of band gaps and the crossing of different modes. The distinctive dispersion properties of gyroids, as highlighted in our results, hold great promise for further research aimed at improving the transmission capabilities of magnonic 3D devices.

\section{Conclusions}

To summarize the research described in this chapter, it is evident that ferromagnetic gyroidal nanostructures exhibit interesting properties with significant implications for the field of 3D magnonics. The presented studies, both static and dynamic magnetization based, have shown that factors such as non-trivial shape anisotropy, chirality, and inhomogeneous demagnetization fields within gyroidal structures are determined by the specific crystallography, and can lead to the creation of multiple low-energy state magnetization textures, SW mode localization, and controllable SW propagation. Though preliminary, these investigations have begun to unveil the substantial potential inherent in these structures.

Further studies of individual nanostructures forming gyroids, such as unit cells, vertices, and struts, are crucial for a comprehensive understanding of the magnetic behavior of gyroids and their interactions with external fields. Much like how micro/nano texturing in metamaterials influences their macroscopic effective properties, these detailed investigations are crucial for elucidating the intricate relationships between gyroid geometry, chirality, and their effective magnetic properties, particularly at microwave frequencies. This is evidenced by the study of resonance frequencies in gyroid samples under rotational field manipulation, demonstrating the significant influence of geometric anisotropy on the FMR signal strength and the structure's effective properties. Gyroids, with their inherently chiral structure, nanoscale unit cell dimensions, and exchange-length bond sizes, show promise for 3D spin-ice systems, exhibiting intriguing collective SW properties such as band gaps and magnon mode hybridizations. However, further experimental studies are essential to validate these properties and demonstrate their practical applications. 
Thus, the results presented in this chapter establish our current understanding of ferromagnetic gyroidal nanostructures and pave the way for their further investigation. In conclusion, research on ferromagnetic gyroidal nanostructures reveals intriguing properties and novel phenomena, underscoring their potential to advance the fields of 3D magnonics and spintronics.

\section*{Acknowledgments}\index{acknowledgments}
The research leading to these results was funded by the National Science Centre of Poland, Projects No.~UMO-2020/39/I/ST3/02413 and No.~UMO-2023/49/N/ST3/03032.

\begin{appendix}
\chapter{Appendix}

\section{Micromagnetic Simulations \label{Sec:sim}}

In Sec.~\ref{sec:fmr}, for the numerical investigation, we used \textit{tetmag} -- the GPU-accelerated open-source FEM micromagnetic solver~\cite{hertel_tetmag_2023}. We exploit its ability to solve magnetostatic open boundary problems in large-scale micromagnetic simulations via a hybrid finite-element/boundary-element formalism~\cite{Hertel2019Large-scale-matrices}. In this analysis, however, we do not assume periodic boundary conditions. All \textit{tetmag}-based micromagnetic simulations were performed in two steps, where the first one was to compute the stable magnetic configuration at a given field value. After the magnetization relaxation, we conducted simulations of the ferromagnetic resonance using a dedicated frequency domain algorithm~\cite{DAquino2023MicromagneticSystems} based on a formulation proposed by~\cite{daquino_novel_2009}.

To investigate the SW modes within the gyroid structure described in Secs.~\ref{sec:localization} and \ref{sec:disp}, we utilized the capabilities of Comsol Multiphysics software. It employs the finite element method (FEM) to solve complex systems of coupled partial differential equations. This includes both the Landau-Lifshitz-Gilbert (LLG) equation and the Maxwell equations under the magnetostatic approximation. In these simulations, each magnetic moment within the unit cells of the gyroid is represented as a normalized unit vector denoted by $\textbf{m}=\textbf{M}/M_\text{S}$, where $\textbf{M}$ is the magnetization distribution function in space and time, and $M_\text{S}$ is the saturation magnetization of the ferromagnetic material.

The core of our approach is solving the LLG equation in its explicit form:
\begin{equation}
\frac{d\textbf{m}}{dt}=\gamma\frac{1}{1+\alpha^2}\left(\textbf{m}\times \textbf{B}_\text{eff}+\alpha\left(\textbf{m}\times\left(\textbf{m}\times \textbf{B}_\text{eff}\right)\right)\right),
\label{Eq:LLGE}
\end{equation}
where $d\textbf{m}/dt$ is the time evolution of the reduced magnetization, $\gamma$ is the gyromagnetic ratio, and $\alpha$ represents the dimensionless damping coefficient. The effective magnetic flux density field, $\textbf{B}_\mathrm{eff}$, integrates the externally applied field, $\textbf{B}_\mathrm{ext}$, with the magnetostatic demagnetizing field, $\textbf{B}_\mathrm{d}$, and the Heisenberg exchange field, $\textbf{B}_\mathrm{exch}$:
\begin{equation}
\textbf{B}_\text{eff}=\textbf{B}_\text{ext}+\textbf{B}_\text{d}+\textbf{B}_\text{exch}.
\end{equation}
The demagnetizing field strength, $\textbf{H}_\text{d}$ (equivalent to $\textbf{B}_\mathrm{d}/\mu_{\text{0}}$, where $\mu_{\text{0}}$ is the vacuum permeability), is governed by Ampère's law, and is derived from the magnetic scalar potential gradient, $U_\text{m}$:
\begin{equation}
\textbf{H}_\text{d}=-\nabla U_\text{m},
\label{h_demag}
\end{equation}
which further evolves within the magnetic body as:
\begin{equation}
\nabla^2 U_\text{m}=\nabla\cdot\textbf{M}.
\label{Eq:nab2}
\end{equation}
In our Comsol implementation, we addressed the eigenproblem derived from Eqs.~\ref{Eq:LLGE}-\ref{Eq:nab2}. Assuming complete magnetization saturation by the magnetic field, we adopted a linear approximation to decompose the magnetization vector into static and dynamic components $\textbf{m}(\textbf{r},t) = m_i \hat{i} + \delta \textbf{m}(\textbf{r},t)\;\forall\;(\delta \textbf{m}\perp\hat{i})$, disregarding all nonlinear terms in the dynamic magnetization $\delta\textbf{m}(\textbf{r},t)$. This method aligns with foundational research such as Refs.~\cite{Mruczkiewicz2013StandingCrystals,Rychy2018SpinRegime}, which provides a comprehensive understanding of the underlying physical principles.

The Bloch-Floquet boundary conditions, applied to parallel surfaces on the opposite sides of the unit cell, are defined as:
\begin{equation}
    \delta\textbf{m}_\text{dst}=\delta\textbf{m}_\text{src}e^{-ik(r_\text{dst}-r_\text{src})}
\end{equation}
where $k$ represents the wavenumber, $\delta \textbf{m}$ is the normalized dynamic component of the magnetization vector at both sides of the unit cell: the destination (dst) and the source (src), $r$ denotes the spatial coordinates of the boundaries where the boundary conditions are applied, and $i$ is the imaginary unit. For a unit cell encompassing periodicity, these boundary conditions are applied to its corresponding parallel faces (as illustrated in Fig.~\ref{Fig:gyr_sizes}). By sweeping the wavenumber parametrically, eigenfrequencies are calculated at each interval, resulting in wavenumber versus frequency plots that reveal the dispersion curves for the periodic structure~\cite{Hakoda2018floquet, Collet2011floquet}. These curves characteristically show periodicity in relation to the wavenumber, with a repeating pattern every $2\pi/L$.

For the distant planes parallel to the axes of the examined gyroid layer, Dirichlet boundary conditions are implemented to nullify the scalar magnetic potential, $U_\text{m}$. These first-type boundary conditions specify the precise value a variable must take at the boundary during the PDE solution process. Here, we set $U_\text{m}|_\text{src}=U_\text{m}|_\text{dst}=0$. To ensure the simulation's physical fidelity and convergence, these conditions are placed sufficiently far from the specimen. In our Comsol simulations, the cell's height was configured to be 40 times the thickness of the gyroid layer.

\section{Broadband FMR experiment \label{Sec:exp}}

By adjusting the position of the sample on the CPW, as described in Sec.~\ref{sec:fmr}, apart from gyroidal structures, we observed a higher frequency signal specifically associated with the homogeneous Ni layer at one edge of the sample. It was confirmed by a good match between the detected signal and the theoretical prediction from the Kittel formula for the resonance frequency:
\begin{equation}
   f = \frac{\gamma}{2\pi} \sqrt{B_\text{ext} (B_\text{ext} + \mu_0 M_\text{s})}.
   \label{eq:kittel}
\end{equation}

In field-swept broadband FMR experiments at a fixed microwave field frequency, the FWHM of the absorption linewidth follows $\mu_0\Delta H = 4\pi\alpha f/\gamma$. This relationship holds when the magnetization vector aligns with the applied magnetic field, either in-plane or perpendicular. The linewidth scales linearly with frequency, with the slope determined by the Gilbert damping parameter $\alpha$. Besides this, empirical data show an additional frequency-independent term:
\begin{equation}
   \Delta H = \Delta H_0 + \frac{4\pi\alpha}{\mu_0\gamma}f,
   \label{eq:deltaH}
\end{equation}
where $\Delta H_0$ represents inhomogeneous contributions, adding to the overall linewidth observed in the experiments.

\end{appendix}

\cleardoublepage

\bibliography{bibliography}

\printindex

\end{document}